\pdfoutput=1
\RequirePackage{ifpdf}
\ifpdf % We~are running pdfTeX in pdf mode
\documentclass[pdftex]{sigma}
\else
\documentclass{sigma}
\fi

\usepackage{mathrsfs}

\numberwithin{equation}{section}

\newtheorem{Theorem}{Theorem}[section]
\newtheorem*{Theorem*}{Theorem}
\newtheorem{Corollary}[Theorem]{Corollary}
\newtheorem{Lemma}[Theorem]{Lemma}
\newtheorem{Proposition}[Theorem]{Proposition}
 { \theoremstyle{definition}
\newtheorem{Definition}[Theorem]{Definition}
\newtheorem{Notation}[Theorem]{Notation}

\newtheorem{Example}[Theorem]{Example}
\newtheorem{Remark}[Theorem]{Remark} }

\begin{document}
\allowdisplaybreaks

\newcommand{\arXivNumber}{2203.09296}

\renewcommand{\thefootnote}{}

\renewcommand{\PaperNumber}{044}

\FirstPageHeading

\ShortArticleName{Field Calculus: Quantum and Statistical Field Theory without the Feynman Diagrams}

\ArticleName{Field Calculus: Quantum and Statistical Field Theory\\ without the Feynman Diagrams\footnote{This paper is a~contribution to the Special Issue on Non-Commutative Algebra, Probability and Analysis in Action. The~full collection is available at \href{https://www.emis.de/journals/SIGMA/non-commutative-probability.html}{https://www.emis.de/journals/SIGMA/non-commutative-probability.html}}}

\Author{John E.~GOUGH}

\AuthorNameForHeading{J.E.~Gough}

\Address{Department of Physics, Aberystwyth University, SY23 3BZ, Wales, UK}
\Email{\href{mailto:jug@aber.ac.uk}{jug@aber.ac.uk}}
\URLaddress{\url{https://www.aber.ac.uk/en/maths/staff-profiles/listing/profile/jug/}}

\ArticleDates{Received March 18, 2022, in final form June 12, 2022; Published online June 14, 2022}

\Abstract{For a given base space $M$ (spacetime), we consider the Guichardet space over the Guichardet space over $M$. Here we develop a ``field calculus'' based on the Guichardet integral. This is the natural setting in which to describe Green function relations for Boson systems. Here we can follow the suggestion of Schwinger and develop a differential (local field) approach rather than the integral one pioneered by Feynman. This is helped by a~DEFG (Dyson--Einstein--Feynman--Guichardet) shorthand which greatly simplifies expressions. This gives a convenient framework for the formal approach of Schwinger and Tomonaga as opposed to Feynman diagrams. The Dyson--Schwinger is recast in this language with the help of bosonic creation/annihilation operators. We also give the combinatorial approach to tree-expansions.}


\Classification{81T18; 05C75; 81S25}

\renewcommand{\thefootnote}{\arabic{footnote}}
\setcounter{footnote}{0}

\section{On three levels}

{\bf Introduction.}
Feynman introduced his diagrammatic approach as a computational technique for quantum electrodynamics~\cite{Schweber} which bypassed quantum field theory. The diagrams are the terms in the perturbative expansion~\cite{Feynman} and their importance is that they allow a systematic approximation starting with the lowest order diagrams. The diagrammatic approach has since found widespread application in the development of quantum field theory \cite{QFT_Diagrams-2,QFT_Diagrams-1,QFT_Diagrams-3}. However, the diagrams are the trees that make up the wood of quantum field theory, and their dominance means that we often cannot see the underlying combinatorial structure. The Feynman approach places emphasis on the particle description at the expense of the local fields themselves. Schwinger considered his approach as the \textit{differential} version of Feynman's \textit{integral} approach, and that the former was mathematically more flexible~\cite{Schwinger}.
Here we build on ideas presented in~\cite{GoughKupsch} to develop a~calculus built around the Guichardet's representation of Fock space~\cite{Guichardet} which is an alternate way to describe quantum field theory which avoids having to draw diagrams and instead uses a~combinatorial identities.

{\bf Guichardet spaces.}
Let $(E, \mathcal{E}, \nu )$ be a $\sigma$-finite measure space. We will be interested in the following collections:
\begin{itemize}\itemsep=0pt\samepage
\item the power set $\mathscr{P} (E)$ -- the collection of all finite sets;
\item the multisets $\mathscr{M} (E)$ -- the collection of all multisets.
\end{itemize}
Recall that a multiset is a collection of elements where repetition is allowed: sets have no repeated elements!

Let $G\colon \mathscr{G} (E) \mapsto \mathbb{C}$. For $X= \{ x_1 ,\dots, x_n \} \in \mathscr{M}(E)$, we may write $G(X) \equiv G_n(x_1 ,\dots, x_n)$ where we now think of $G_n$ as a completely symmetric function of $n$-variables. We shall say that~$G$ is integrable if it is measurable and $\sum_{n \ge 1} \frac{1}{n!} \int_{E^n} | G_n( x_1 ,\dots, x_n) | \mu[{\rm d}x_1] \cdots \mu [{\rm d}x_n ] < \infty$. For integrable~$G$, we then set
\begin{gather}
\int_{\mathscr{M} (E)} G(X) \mu [{\rm d}X] = G(\varnothing ) + \sum_{n \ge 1} \frac{1}{n!} \int_{E^n} G_n( x_1 ,\dots, x_n) \mu[{\rm d}x_1] \cdots \mu [{\rm d}x_n ].
\label{eq:G-integral}
\end{gather}
In this way, the measure $\mu [{\rm d}x ]$ on $E$ induces a measure $\mu [{\rm d}X ]$ on $\mathscr{M} (E)$ which is known as the Guichardet measure~\cite{Guichardet}.

\begin{Example}[atomic measure]
If we take $\mu $ to be pure point, that is $\mu [{\rm d}x] = \sum_{a \in A} m(a) \delta_a [{\rm d}x]$ where the support $A$ is countable, then
the induced measure assigns a weight $1$ to the empty set, $m(a_1) \cdots m(a_n)$ to a multiset $\{ a_1,\dots, a_n \}$ with all elements in $A$, and zero to everything else. The expression~(\ref{eq:G-integral}) becomes
 \begin{gather*}
 G(\varnothing ) + \sum_{n \ge 1} \frac{1}{n!} \sum_{a_1 \in A } \cdots \sum_{a_n \in A}
 G_n( a_1 ,\dots, a_n) m(a_1) \cdots m(a_n).
\end{gather*}
\end{Example}

\begin{Example}[non-atomic measure]
If we take $\mu $ to be non-atomic, then the diagonal elements can be safely ignored. In this case, we may expressly drop the repetitions and use sets rather than multisets:
\begin{gather*}
\int_{\mathscr{M} (E)} G(X) \mu [{\rm d}X] \equiv \int_{\mathscr{P} (E)} G(X) \mu [{\rm d}X].
\end{gather*}
\end{Example}

The space $L^2 ( \mathscr{M} (E) , \mu [{\rm d}X])$ is called the Guichardet space over $(E , \mathcal{E} , \mu [{\rm d}x] )$. The inner product is given by
\begin{gather*}
\int_{\mathscr{M} (E)} \Psi(X)^\ast \Phi (X) \mu [{\rm d}X] \\
\qquad{} = \Psi(\varnothing )^\ast \Phi (\varnothing)+ \sum_{n \ge 1} \frac{1}{n!} \int_{E^n} \Psi_n ( x_1 ,\dots, x_n)^\ast \Phi ( x_1 ,\dots, x_n) \mu[{\rm d}x_1] \cdots \mu [{\rm d}x_n ].
\end{gather*}
This is the Fock space over $L^2(E , \mathcal{E} , \mu [{\rm d}x] )$.

\begin{Proposition}
Let $\mu [{\rm d}x] = \mu_c [{\rm d}x] + \mu_a [{\rm d}x]$ be the decomposition of a measure on $E$ $($with $A$ the support of~$\mu_a)$ into non-atomic and atomic parts, then
\begin{gather}
L^2 ( \mathscr{M} (E) , \mu [{\rm d}X]) \cong L^2 ( \mathscr{P} (E) , \mu_c [{\rm d}X]) \otimes L^2 ( \mathscr{M} (A) , \mu_a [{\rm d}X]) .
\label{eq:cong}
\end{gather}
\end{Proposition}
Effectively, we have $L^2(E , \mathcal{E} , \mu [{\rm d}x] ) \cong L^2(E , \mathcal{E} , \mu_c [{\rm d}x] ) \oplus L^2(E , \mathcal{A} , \mu_a [{\rm d}x] )$
and use the fact that the Fock space over a direct sum of Hilbert spaces is naturally isomorphic to the tensor product of the Fock spaces over the individual spaces, see for instance~\cite[Chapter~II]{Partha}.

The first factor in (\ref{eq:cong}) is the Fock space for a field while the second can be thought of as the Hilbert space of an assembly of harmonic oscillators (one for every $a \in A$).

\begin{Remark}\label{rem:anon_ref}
The induced measure $\mu [{\rm d}X]$ on the multisets will always have an atomic component. Even in the case where $\mu [{\rm d}x]$ is non-atomic, the measure $\mu [{\rm d}X]$ will still have an atom at~$\varnothing$.
\end{Remark}

\begin{Proposition}\label{prop:ca}
Let $\mu [{\rm d}x] = \mu_c [{\rm d}x] + \mu_a [{\rm d}x]$ be the decomposition of a measure on $E$ $($with $A$ the support of $\mu_a)$
into non-atomic and atomic parts, then
\begin{gather*}
\int_{\mathscr{M} (E)} G(X) \mu [{\rm d}X] = \int_{\mathscr{P} (E)} \mu_c [{\rm d}X] \int_{\mathscr{M} (A)} \mu_a [{\rm d}X] G( X \cup Y ) ,
\end{gather*}
where $X\cup Y$ is the union of multisets $($adding degeneracies where they occur$)$.
\end{Proposition}
\begin{proof} The integral $\int_{\mathscr{M} (E)} G(X) \mu [{\rm d}X]$ may be written as
\begin{gather*}
\sum_{r=0}^\infty \frac{1}{r!} \int_{E^r} \sum_{n=0}^r \binom{r}{n} G( x_1 ,\dots, x_r )
\mu_c [{\rm d}x_1] \cdots \mu_c [{\rm d}x_n] \mu_a [{\rm d}x_{m+1}] \cdots \mu_a [{\rm d}x_r]
\end{gather*}
and introducing $m = r-n$ we may rewrite this as
\begin{gather*}
\sum_{n=0}^\infty \sum_{m=0}^\infty \frac{1}{n!m!} \int_{E^{n+m}} G( x_1 ,\dots, x_n , y_1,\dots, y_m )
\mu_c [{\rm d}x_1] \cdots \mu_c [{\rm d}x_m] \mu_a [{\rm d}y_1 ] \cdots \mu_a [{\rm d}y_n ]
\end{gather*}
yielding the result.
\end{proof}

As a corollary, we may absorb the atomic contribution and obtain the identity
\begin{gather*}
\int_{\mathscr{M} (E)} G(X) \mu [{\rm d}X] = \int_{\mathscr{P} (E)} \tilde{G} (X) \mu_c [{\rm d}X] ,
\end{gather*}
where we introduce the ``renormalized'' function
\begin{gather*}
\tilde{G} (X) \triangleq \int_{\mathscr{M} (A)} G(X+Y ) \mu_a [{\rm d}Y] .
\end{gather*}

\section{Green's functions}
Let $M$ be Minkowski spacetime with $dx$ denoting the 4-volume element. This will be the level~1 description of fields. Specifically, a field $\varphi$ over $M$ is a complex-valued function on $M$ and its value at a point $x\in M$ is denoted either as $\varphi (x)$ or $\varphi_x$.

Let $G=(G_n)_n$ be a family of completely symmetric functions of an indefinite number of spacetime variables: that is, for each $n=0,1,2,\dots$ we have $G_n\colon \times^n M \mapsto \mathbb{R}$ with $G_n (x_1,\dots, x_n)$ invariant under arbitrary permutation of the arguments. The family may then be identified as a function $G\colon \mathscr{M}(M) \mapsto \mathbb{C}$ by
\begin{gather*}
 G(X) \equiv G_n (x_{1},\dots ,x_{n}) ,%\label{eq:G}
\end{gather*}
whenever $X= \{x_{1},\dots ,x_{n}\}$ is a multiset. This includes the value $G_{0}$ which is assigned to the empty set: $G_0 \equiv G(\varnothing)$.

As the measure ${\rm d}x$ is non-atomic, in standard configuration there will be no degeneracy allowing us to restrict attention to just~$\mathscr{P} (M)$. The power set of~$M$ will be our level~2 description.

The measure ${\rm d}x$ at level~1 induces the Guichardet measure ${\rm d}X$ \cite{Guichardet}, at level~2 and we have specifically
\begin{gather*}
\int_{\mathscr{P}(M)} G(X){\rm d}X \triangleq \sum_{n\geq 0}
\frac{1}{n!}\int_{\times^n M} G_n(x_{1},\dots ,x_{n})\,{\rm d}x_{1}\cdots {\rm d}x_{n}.
\end{gather*}
The Guichardet measure was extensively used in the kernel approach to quantum stochastic processes by Hans Maassen~\cite{Maassen85} and Paul-Andre Meyer \cite{QP4P}. The extension to space follows from an important remark due to Joachim Kupsch~\cite{QP4P} and, as Meyer remarks, the results may be extended to a Luzin space $M$ with a non-atomic measure.

Our next step is to apply the Guichardet construction to level~2, where we encountered the measure space $( \mathscr{P}(M) , {\rm d}X )$, and go up another level. Here we will encounter $\mathscr{M} (\mathscr{P}(M))$ but as before this is too big and we can reduce. However, following Remark~\ref{rem:anon_ref}, we have that ${\rm d}X$ has a single atom at $X = \varnothing$. Effectively, we should restrict to $\mathscr{P}_\varnothing (\mathscr{P}(M))$ which we take to be the multisets of $\mathscr{P}(M)$ where only the empty set $\varnothing$ is allowed to repeat.
The basic elements are now $\mathcal{X} = \{ X_1 ,\dots,X_n\}$ which are collections of subsets none of which are equal unless they are the empty set. The functions $G (\mathcal{X}) \equiv G(X_1,\dots, X_n)$ are again completely symmetric in there arguments and the induced measure ${\rm d}\mathcal{X}$ is given by
\begin{gather*}
\int_{\mathscr{M} (\mathscr{P}(M))} G(\mathcal{X}){\rm d}\mathcal{X} \triangleq \sum_{n\geq 0}
\frac{1}{n!}\int_{\times^n \mathscr{P} (M)} G_n(X_{1},\dots ,X_{n})\, {\rm d}X_{1}\cdots {\rm d}X_{n},
\end{gather*}
so that we obtain the measure space $( \mathscr{P}_\varnothing (\mathscr{P}(M)) , {\rm d}\mathcal{X})$. We have the equivalence
\begin{gather*}
\int_{\mathscr{M} (\mathscr{P}(M))} G(\mathcal{X}){\rm d}\mathcal{X} \equiv
\int_{\mathscr{P}_\varnothing (\mathscr{P}(M))} G(\mathcal{X})\,{\rm d}\mathcal{X} .
\end{gather*}

\begin{center}
		\begin{tabular}{c|l|l|l}
			Level & Framework & Elements & Measure\\ \hline\hline
			1 & spacetime & $x$ & ${\rm d}x$ \\
			2 & subsets of spacetime & $X = \{ x_ 1 ,\dots, x_n \}$ & ${\rm d}X$ \\
			3 & subsets of subsets of spacetime & $ \mathcal{X} = \{ X_1 ,\dots, X_n\} $ & ${\rm d} \mathcal{X}$
		\end{tabular}
\end{center}

Why would we want to go up another level? It turns out that we naturally encounter subsets of subsets of spacetime events when dealing with fields, and this level will turn out to be highly convenient when we try to describe combinatorial features.

For $X \in \mathscr{P}(M)$, we will denote the number of elements as $\# X$. An equation of the form $X+Y=Z$ signifies that $Z$ is the union of disjoint sets $X$ and~$Y$. A \textit{decomposition} of $X$ is an ordered sequence $(X_1 ,\dots,X_n)$ of disjoint sets (some of which may be empty!) whose union is~$X$.

\begin{Definition}A \textit{partition} of $X\in \mathscr{P}(M)$ is a set $\mathcal{X} \in \mathscr{P}(\mathscr{P}(M)) $ of (unordered) disjoint non-empty subsets (which we then refer to as parts) whose union is $X$. We also introduce the more general concept of a \textit{division} of $X$ which allows some of the parts to be empty: divisions naturally belong to $\mathscr{P}_\varnothing (\mathscr{P}(M)) $. The number of subsets making up a division/partition $\mathcal{X}$ is denoted as $N(\mathcal{X})$. We will write $\mathrm{Div}(X)$ and $\mathrm{Part}(X)$ for the collection of all divisions and partitions of~$X$ respectively. We denote by $\mathrm{Pair} (X)$ the collection of all pair partitions.
\end{Definition}

For definiteness, a division of $X$ will be a set $\mathcal{X} = \{ X_1 ,\dots, X_n \}$ where $X_j \cap X_k = \varnothing$ for all $j \neq k$, and $X_1 + \cdots +X_n =X$. In this case $N(\mathcal{X} ) = n$. It will be a partition if we furthermore have $\# X_k \ge 1$ for each $k$. In the case of pair partitions, we have the additional constraint that $\# X_k =2$ for each $k$ (so $\# X$ must be even).

\begin{Notation}[Einstein--Guichardet convention]
We use lowercase letters to denote functions on $M$: a repeated spacetime index $x$ implies integration with respect to ${\rm d}x$ over~$M$:
\begin{gather*}
f^xg^x \triangleq \int_M f(x) g(x) \,{\rm d}x .
\end{gather*}
We use uppercase letters to denote functions on $\mathscr{P}(M)$: a repeated subset index implies integration with respect to ${\rm d}X$ over $\mathscr{P}(M)$:
\begin{gather*}
F^X G^X \triangleq \int_{\mathscr{P} (M)} F(X) G(X)\, {\rm d}X .
\end{gather*}
Finally, we use uppercase calligraphic letters to denote functions on $\mathscr{P}_\varnothing (\mathscr{P}(M))$: a repeated subset index implies integration with respect to ${\rm d}\mathcal{X}$ over $\mathscr{P}_\varnothing (\mathscr{P}(M))$:
\begin{gather*}
F^\mathcal{X} G^\mathcal{X} \triangleq \int_{\mathscr{P}_\varnothing (\mathscr{P}(M))} F(\mathcal{X}) G(\mathcal{X})\,{\rm d}\mathcal{X} .
\end{gather*}
\end{Notation}
Note that we place arguments as either subscripts or superscripts. For functions on $M$, the idea is that the fields are covariant and so carry a subscript: $\varphi (x) \equiv \varphi_x$. Dual to these we will have sources $j(x) \equiv j^x$ which carry a contravariant index. The duality is the $\langle j,\varphi \rangle = j^x \varphi_x$. We will take $\Phi$ to be a suitable class of fields, and~$\mathfrak{J}$ to be a suitable class of sources.

\begin{Notation}
A further shorthand is the following: let $f$ be measurable on $M$ and $X\in \mathscr{P} (M)$ then write
\begin{gather*}
f^{X}\triangleq \prod_{x\in X}f(x). %\label{eq:f^X}%\label{eq:Guichardet}
\end{gather*}
Likewise, we use $F^\mathcal{X} = \prod_{X\in \mathcal{X}} F^X$.
\end{Notation}

As a first foray into the Einstein--Guichardet convention, let us show the following result

\begin{Proposition}\label{prop:exp}
We have the following identities:
\begin{gather*}
f^X g^X \equiv {\rm e}^{f^x g^x} , \qquad F^\mathcal{X} G^\mathcal{X} \equiv {\rm e}^{F^X G^X} .
\end{gather*}
\end{Proposition}
\begin{proof}
The right hand side in the first identity reads as follows
\begin{align*}
\int_{\mathscr{P}(M)} f^X g^X \, {\rm d}X &= \sum_{n\geq 0}
\frac{1}{n!}\int_{\times^n M} f(x_{1})\cdots f(x_{n}) g(x_1) \cdots g(x_n)\,{\rm d}x_{1}\cdots {\rm d}x_{n} \nonumber \\
&=
\sum_{n\geq 0}
\frac{1}{n!} \bigg[ \int_{ M} f(x) g(x)\,{\rm d}x \bigg] ^n = \exp \int_M f(x) g(x) \, {\rm d}x .
\end{align*}
The second identity is just the level 3 analogue.
\end{proof}

With the convention $f^\mathcal{X} = \prod_{X \in \mathcal{X}} \prod_{x \in X} f(x)$, we have
\begin{gather*}
f^\mathcal{X} g^\mathcal{X} \equiv {\rm e}^{{\rm e}^{f^xg^x}} .
\end{gather*}
Note that the exponential of an exponential arises naturally as generating functional associated with Poisson point process distributions.

The space $L^2 (\mathscr{P} (M), {\rm d}X)$ is, in fact, then the Fock space over $L^2( M ,{\rm d}x)$. Given a wave-function $\Psi$ for an indefinite number of boson particles, we may write it as function $\Psi\colon X \mapsto \Psi (X) $, and we have the overlaps
\begin{gather*}
\langle \Phi | \Psi \rangle \triangleq \int \Phi (X) ^\ast \Psi (X)\, {\rm d}X \equiv (\Phi^\ast)^X \Psi^X .
\end{gather*}
In particular, the function $\exp^f$ defined by $\exp^f (X) = f^X$ is an exponential vector and from Proposition~\ref{prop:exp} we have that $\langle \exp^f | \exp ^g \rangle = {\rm e}^{ \langle f |g \rangle}.$

Moving up a level, we see that $L^2 (\mathscr{P}_\varnothing (\mathscr{P}(M)), {\rm d}\mathcal{X} )$ is the Fock space over $L^2 (\mathscr{P} (M), {\rm d}X)$- that is, the Fock space over the Fock space over $L^2( M ,{\rm d}x)$.

\subsection{Generating functionals and the Wick product}

\begin{Definition}
Let $G$ be a measurable function on $\mathscr{P} (M)$ then we define its associated generating functional, for test functions $j $ belonging to some suitable class $ \mathfrak{J}$, as
$Z_{G}(j)= G_X j^{X} $ or, in longhand
\begin{gather*}
 Z_{G} (j) \triangleq \sum_{n\geq 0}%
\frac{1}{n!}\int G(x_{1},\dots ,x_{n}) j(x_{1})\cdots
j(x_{n})\,{\rm d}x_{1}\cdots {\rm d}x_{n} .
 %\label{eq:Z_G_shorthand}
\end{gather*}
\end{Definition}

\begin{Definition}
The Wick product of functions on $\mathscr{P} (M)$ is defined to be
\begin{gather*}
F\diamond G(X)\triangleq \sum_{X_{1}+X_{2}=X}F(X_1) G(X_2). %\label{eq:wick_prod}
\end{gather*}
Here the sum is over all decompositions of the set $X$ into ordered pairs $(X_{1},X_{2})$ whose union is~$X$.
\end{Definition}

The importance of the Wick product is revealed in the next result.

\begin{Proposition}
We have $ Z_{F\diamond G}=Z_{F}Z_{G}$, where the $F\diamond G$ is the Wick product.
\end{Proposition}
\begin{proof}
In our Einstein--Guichardet notation, we have that $Z_F (j) Z_G (j) = F_{X_1} j^{X_1} G_{X_2} j^{X_2}= F_{X_1} G_{X_2} j^{X_1 +X_2}$ which we rewrite as $(F \diamond G)_X j^X $.
\end{proof}

To appreciate the compactness of the notation, let us redo the proof longhand. We have that $\int F(Y)G(Z)j^{Y+Z}\,{\rm d}Y{\rm d}Z$ equals
\begin{gather*}
 \sum_{n_{1},n_{2}\geq 0}
\frac{1}{n_{1}!n_{2}!}\int F(y_{1},\dots ,y_{n_{1}})G(z_{1},\dots
,z_{n_{2}}) \\
 \qquad\qquad \qquad {}\times j(y_{1})\cdots j(y_{n_{1}})j(z_{1})\cdots
j(z_{n_{2}})\,{\rm d}y_{1}\cdots {\rm d}y_{n_{1}}{\rm d}z_{1}\cdots {\rm d}z_{n_{2}} \\
 \qquad{} \equiv \sum_{n\geq 0}\frac{1}{n!}\int F\diamond G(x_{1},\dots
,x_{n}) j(x_{1})\cdots j(x_{n})\,{\rm d}x_{1}\cdots {\rm d}x_{n}
\end{gather*}
and comparing powers of $j$ we get $F\diamond G(X)=\sum_{Y+ Z=X}F(Y)G(Z)$.

The Wick product is a symmetric associative product with general form
\begin{gather*}
G_{1}\diamond \cdots \diamond G_{n}(X)=\sum_{X_{1}+\cdots+X_{n}=X}G_{1}(X_{1})\cdots G_{n}(X_{n}),
\end{gather*}
where the sum is now over all decomposition of $X$ into $n$ ordered disjoint subsets.

Some care is needed when the Green's functions are the same. Denoting the $n$th Wick power of $F$ by $F^{\diamond n}=F\diamond \cdots \diamond F$ ($n$ tines), we have
\begin{gather*}
F^{\diamond n}(X) =\sum_{X_{1}+\cdots +X_{n}=X}F(X_{1})\cdots F(X_{n}) .
\end{gather*}

This leads to the following Wick product of exponential vectors: $\exp (f)\diamond \exp (g)\equiv \exp (f+g)$. The next result is due to Hans Maassen \cite{Maassen85}.

\begin{Lemma}[the $\Sigma \hspace{-0.12in} \int $ lemma]\label{lem:Sum-Int}
For $F $ a measurable function on $\times ^{p}\mathscr{P}(M)$ we have the identity
\begin{gather*}
\int F ( X_{1},\dots ,X_{p} ) \,{\rm d}X_{1}\cdots {\rm d}X_{p}\equiv \int \sum_{X_{1}+\cdots +X_{p}=X}F ( X_{1},\dots ,X_{p} )\, {\rm d}X.
%\label{eq:Sum-Int}
\end{gather*}
\end{Lemma}

\begin{proof}If we write both of these expressions out longhand, then the left hand side picks up the factors $\frac{1}{n_{1}!\cdots n_{p}!}$ where $\# X_{k}=n_{k}$.
On the right hand side we get $\frac{1}{n!}$ where $\#X=n$. Both expressions are multiple integrals with respect to either ${\rm d}X_{1}\cdots {\rm d}X_{p}$ or ${\rm d}X$
with $X_{1}+\cdots +X_{p}=X$ however on the right hand side we obtain an additional factor $\binom{n}{n_{1},\dots ,n_{p}}$ giving the number of
decompositions of~$X$ with $n_{k}$ elements in the $k$th subset. This accounts precisely the combinatorial factors so both sides are equal.
\end{proof}

Note that this is also related to the identity in Proposition~\ref{prop:ca}.

\subsection{The composition formula}
Suppose we are given a generating functional $K_X j^X$ then its exponential will also take the form of a generating functional, say $G_X j^X$. We have
\begin{gather*}
G_X j^X ={\rm e}^{K_X j^X} = K_\mathcal{X} j^\mathcal{X},
\end{gather*}
where $j^\mathcal{X} = \prod_{X \in \mathcal{X}} \prod_{x \in X} j^x$, that is, the product of all~$j^x$ where~$x$ is one of the spacetime events making up $\mathcal{X}$. We see that
\begin{gather*}
G_X = \sum_{\mathcal{X} \in \mathrm{Div}(X)} K_{\mathcal{X}}
\equiv \sum_{n}\frac{1}{n!}(K ^{\diamond n})_X.
\end{gather*}
If we furthermore assume that $K_\varnothing =0$, then the nontrivial divisions come from partitions and so
\begin{gather*}
G_X = \sum_{\mathcal{X} \in \mathrm{Part}(X)} K_{\mathcal{X}}.
\end{gather*}
For instance, $G_\varnothing = 1$, $G_{ \{ x_1 \}} = K_{\{ x_1 \} }$, $G_{ \{ x_1 , x_2 \}} = K_{\{ x_1 , x_2 \} } + K_{\{ x_1 \} }
K_{ \{ x_2 \} }$, etc.

\begin{Theorem}
Let $h$ be a function with Maclaurin series $h(z)=\sum_{n}\frac{1}{n!}h_{n}z^{n}$ and $F_Xj^X$ a~generating functional, then
$h(Z_{F}(j)) \equiv H_X j^X$ where
\begin{gather*}
H_X =\sum_{\mathcal{X} \in \mathrm{Div}(X)}h_{N(\mathcal{X} )}F_{\mathcal{X} }. %\label{eq:comp_gen}
\end{gather*}
\end{Theorem}
Here $H_X \equiv \sum_{n}\frac{1}{n!}h_{n}(F^{\diamond n})_X=\sum_{n}h_{n}\sum_{X_{1}+\cdots +X_{n}=X}^{\text{unordered}}F(X_{1})\cdots F(X_{n})$ where we drop the ordering of the elements of the decomposition which absorbs the~$n!$.

We list out some important examples.
\begin{enumerate}\itemsep=0pt
\item \textit{Exponentials} $G_Xj^X = {\rm e}^{K_X j^X}$: $G_X = \sum_{\mathcal{X} \in \mathrm{Div}(X)} K_{\mathcal{X} }$.
\item \textit{Logarithms} $K_Xj^X = \ln G_X j^X$: $K_X = \sum_{\mathcal{X} \in \mathrm{Div}(X)} \mu_{N (\mathcal{X})} G_{\mathcal{X}}$, with $ \mu_n = (-1)^n (n-1)!$.
\item \textit{Inverse} $F_Xj^X = \frac{1}{ G_X j^X}$: $F_X = \sum_{\mathcal{X} \in \mathrm{Div}(X)} \nu_{N (\mathcal{X})} G_{\mathcal{X}}$, with $\nu_n = (-1)^{(n+1)} n! $.
\end{enumerate}

\section{Random fields}
A field $\varphi$ is a complex-valued function on $M$. We will write $\varphi (x)$ as $\varphi_x$ (covariant index) and denote the space of fields as $\Phi$.
We will dual space $\mathfrak{J}$ of source fields and write $j(x) = j^x$ and the duality is written as $j^x \varphi_x$ (contravariant index).

A multi-linear map $T\colon \times ^{n}\mathfrak{J}\mapsto \mathbb{C}$ is called a tensor of covariant rank $n$ and it will be determined by the components
$T_{x_{1}\dots x_{n}}$ such that $T( j_{( 1) },\dots ,j_{( n) }) =T_{x_{1}\dots x_{n}} j_{(1)
}^{x_{1}}\cdots j_{(n) }^{x_{n}}$. We take the dual space to $\mathfrak{J}$ to be $\Phi$ and this is the space in which the fields live. Likewise, we refer to a multilinear map from $\times ^{n}\Phi $ to the complex numbers as a tensor
of contravariant rank~$n$.

\subsection{Random fields and their Green's functions}

A random field $\phi $ is a random variable $\phi_x$ to each point $x\in M$. More exactly, it is determined by its characteristic functional
\begin{gather*}
Z_G (j) \triangleq \mathbb{E}\big[ {\rm e}^{j^x\phi_x }\big]
=\mathbb{E} \big[ j^X \phi_X \big] = j^X G_X
\end{gather*}
for suitable functions $j$ belonging to some class $\mathfrak{J}$. It's moments are the Green's functions $G( x_{1},\dots ,x_{n}) =\mathbb{E}
 [ \phi _{x_{1}}\cdots \phi _{x_{n}} ] $ and we may write this as
\begin{gather*}
G_{X}=\mathbb{E} [ \phi _{X} ] =\left. \frac{\delta Z ( j )
}{\delta j^{X}}\right| _{j=0},
\end{gather*}
where we introduce the shorthand $\phi _{X}=\prod_{x\in X}\phi _{x}$.

The cumulant Green's functions $K_X$ are defined by $K_X jX= \ln G_X j^X$. As $G_\varnothing =1$, we have $K_\varnothing = 0$ and from the composition formula the two types of moments are related by
\begin{gather*}
G_X = \sum_{\mathcal{X} \in \mathrm{Part}(X)} K_{\mathcal{X} }, \qquad
 K_X = \sum_{\mathcal{X} \in \mathrm{Part}(X)} (-1)^{N (\mathcal{X})} N( \mathcal{X} )! G_{\mathcal{X}}.
\end{gather*}

\begin{Definition}A function $A$ on fields is said to be analytic if it takes the form $A(\varphi ) = A^X \varphi_X$, or in longhand
\begin{gather*}
A (\varphi ) = \sum_{n\geq 0}
\frac{1}{n!}\int_{\times^n M} A^{\{ x_{1},\dots ,x_{n}\} } \varphi_{x_1} \cdots \varphi_{x_n} \,{\rm d}x_{1}\cdots {\rm d}x_{n}.
\end{gather*}
\end{Definition}

Let $A$ be an analytic function on $\Phi$, say $A (\varphi ) = a^X \varphi_X$, then the average of is $A(\phi )$ is
\begin{gather*}
\mathbb{E} [ A(\phi ) ] = a^X \mathbb{E} [ \phi_X ] = a^X G_X .
\end{gather*}
This suggests that a useful way to think of the averaging process for fields is as a dual map acting on the sequences $\big(A^X\big)_X$ of coefficients. The moments $(G_X)_X$ act as the dual coefficients and must satisfy normalization, $G_\varnothing =1$, and positivity. We should have, for instance, that for each integer $N$ and for each set of (complex) sources $j_1,\dots , j_N \in \mathfrak{J}$, the property $ 0 \le \mathbb{E} \big[ \big| \sum_{n=1}^N {\rm e}^{j^x_n \phi_x } \big|^2 \big]$ which equates to the following Bochner condition $\sum_{n,m=1}^N G_X (j_n^\ast -j_m)^X \ge 0$.

\begin{Lemma}Let $\mathbb{E} [ \cdot] $ be an expectation with the corresponding Green's
function $G_{X}=\mathbb{E}[ \phi _{X}] $. For analytic
functionals $A( \phi ) =\int A^{X}\phi _{X}{\rm d}X$, $B( \phi ) =\int B^{X}\phi _{X}{\rm d}X$, etc., we have the formula
\begin{gather}
\mathbb{E} [ A ( \phi ) B ( \phi ) \cdots Z (\phi ) ] = ( A\diamond B\diamond \cdots \diamond Z) ^{X} G_X. \label{eq:diamond formula}
\end{gather}
\end{Lemma}

\begin{proof}The expectation in (\ref{eq:diamond formula}) reads as
\begin{align*}
 A^{X_{a}}B^{X_{b}}\cdots Z^{X_{z}}\mathbb{E} [ \phi _{X_{a}}\phi_{X_{b}}\cdots \phi _{X_{z}} ]
& = A^{X_{a}}B^{X_{b}}\cdots Z^{X_{z}}G_{X_{a}+X_{b}+\cdots
+X_{z}}\\
& \equiv \bigg( \sum_{X_{a}+X_{b}+\cdots
+X_{z}=X}A^{X_{a}}B^{X_{b}}\cdots Z^{X_{z}}\bigg) G_{X},
\end{align*}
where we use the $\Sigma \hspace{-0.12in} \int $ lemma.
\end{proof}

\subsection{Gaussian states}

We now describe how to define Gaussian states on $\Phi$.

\begin{Definition}A metric $g\colon \times^2 \mathfrak{J} \mapsto \mathbb{R}$ is an invertible symmetric tensor. We write $g_{xy}$ for the components of the tensor: symmetry means $g_{xy}=g_{yx}$ and invertibility means that there exists a contravariant tensor $g^{-1}\colon \times^2 \Phi \mapsto \mathbb{R}$ with components $g^{xy} $ with $g_{xy} g^{yz} = \delta (x-z)$.
\end{Definition}

It practice, the metric $g$ corresponds to kernel function associated with the free field operator. For further discussion of Gaussian measure for such problems see, for instance, \cite[Chapters~6 an~7]{GlimmJaffe}.

Let $\overline{\phi} \in \Phi$ and $g$ be a metric. The \textit{Gaussian state} with mean $\overline{\phi}$ and covariance $g$ is then determined by the characteristic function
\begin{gather*}
 \mathbb{E}_{\overline{\phi}, g } \big[ {\rm e}^{j^x \varphi_x} \big] = {\rm e}^{j^x \overline{\phi}_x + \frac{1}{2} j^x g_{xy} j^y }.
\end{gather*}

If we set the mean field $\overline{\phi } =0$, then the Gaussian state is completely characterized by the fact that the only
non-vanishing cumulant is now $K_{\{ x,y\} }=g_{xy}$. We will denote this by $\mathbb{E}_g$.
If we now use the identity $G_X= \sum_{\mathcal{X} \in \mathrm{Part}(X)} K_\mathcal{X}$ to obtain the Green's
functions, then we see that all odd moments vanish, while
\begin{gather}
\mathbb{E}_g [ \phi _{x(1) }\cdots \phi _{x(2k) }] =\sum_{\mathrm{Pair}( 2k) }g_{x(p_{1}) x( q_{1}) }\cdots g_{x( p_{k}) x(q_{k}) },
\label{eq:Gaussian field moments}
\end{gather}
where the sum is over all pair partitions of $\{ 1,\dots ,2k\} $. The right-hand side will of course consist of $\frac{(2k) !}{2^{k}k!}$ terms. To this end we introduce some notation. Let $\mathcal{P}\in
\mathrm{Pair}(X) $ be a given pair partition of a subset $X$, say $\mathcal{P}=\{ x_{p(1)},x_{q(1)}),\dots ,(x_{p(k)},x_{q(k)})\} $, then we write $g_{\mathcal{P}}=g_{x_{p(1) }x_{q(1) }}\cdots g_{x_{p(k) }x_{q(k) }}$ in which case the Gaussian moments are
\begin{gather}
G_{X}^{g}\equiv \mathrm{Pair}(g)_X . \label{eq:Gaussian_moments}
\end{gather}
where we introduce
\begin{gather*}
\mathrm{Pair}(g)_X \triangleq \sum_{\mathcal{P}\in \mathrm{Pair}(X) }g_{\mathcal{P}}.
\end{gather*}

\subsection{General states}
Suppose that we have an analytic functional on the fields, say $V (\varphi ) = V^{X}\varphi _{X} $. (If we wish, we can take $V^{x,y} \equiv 0$ as otherwise it would be absorbed into the covariance~$g$. We typically take $V^\varnothing =0$ as this plays no role.) We additionally assume that
\begin{gather*}
\Xi \equiv \mathbb{E}_{g}\big[ {\rm e}^{V ( \phi ) }\big] < \infty .
\end{gather*}

Given a reference Gaussian state with covariance $g$ (a metric!), we then obtain new state according to
\begin{gather}
\mathbb{E} [ A( \phi ) ] =\frac{1} { \Xi } \mathbb{E}_g \big[ A(\varphi ) {\rm e}^{V ( \phi ) } \big] .\label{eq:relative_state}
\end{gather}

A formal approach is to introduce functional integrals. Let $S$ be a functional on $\Phi$, which we call the \textit{action} and consider formal expressions of the form
\begin{gather*}
\mathbb{E} [A(\phi)] = \frac{ \int_\Phi A(\varphi ) {\rm e}^{S(\phi )} \mathscr{D} \varphi} { \int_\Phi {\rm e}^{S(\phi )} \mathscr{D} \varphi}.
%\label{eq:functional_integral}
\end{gather*}

For instance, let us then set
\begin{gather*}
S_0 (\varphi ) = - \frac{1}{2} \varphi_x g^{xy} \varphi_y ,
\end{gather*}
and suppose (by analogy to standard Gaussian integrals) that
\begin{gather*}
 \int_\Phi {\rm e}^{j^x \varphi_x} {\rm e}^{S_0(\varphi )} \mathscr{D} \varphi = {\rm e}^{ \frac{1}{2} j^x g_{xy} j^y }.
\end{gather*}
To be clear, this is all formal! All we have done is to say that the expression ``${\rm e}^{S_0(\phi )} \mathscr{D} \varphi$'' appearing in the functional integral can be made rigorous by interpreting it as a Gaussian measure $\mathbb{P}_g [{\rm d}\varphi ]$ which gives us the well defined Gaussian state $\mathbb{E}_g$ introduced before. The general state is then $ \mathbb{P} [ {\rm d} \varphi ] = {\rm e}^{ V (\varphi ) } \mathbb{P}_g [ {\rm d} \varphi ]$. Formally, this corresponds to the action
\begin{gather*}
S(\varphi ) = S_0 (\varphi ) + V(\varphi ) .
\end{gather*}

\section{Why we need level 3 for random fields!}
In the average (\ref{eq:relative_state}) we encounter the exponential ${\rm e}^{V(\phi )}$. We can use (\ref{eq:diamond formula}) to get $\mathbb{E}_g[ V ( \phi ) ^n] = ( V^{\diamond n}) ^{X}G_{X}^g$, but there is a more natural way to calculate $\mathbb{E}_g\big[ {\rm e}^{V(\phi ) }\big] $. By Proposition~\ref{prop:exp}, we have $ {\rm e}^{V(\phi ) } = {\rm e}^{V^X \phi_X } \equiv V^\mathcal{X} \phi_{\mathcal{X}}$ and the Gaussian expectations yields
\begin{gather*}
\Xi = V^\mathcal{X} \mathbb{E}_g [ \phi_\mathcal{X} ] = \sum_{n=0}^\infty \frac{1}{n!} V^{X_1}\cdots V^{X_n} G^g_{X_1 + \cdots + X_n }.
\end{gather*}
As we have $V^{\varnothing }=0$, the sum may be considered as being over all partitions $\mathcal{X}$ and this will further absorb the $n!$ factor since the partitions are unordered. We obtain
\begin{gather*}
\Xi = \mathrm{Part} (V)^X G^g_{X}
\end{gather*}
where we introduce $\mathrm{Part} (V)^X = \sum_{\mathcal{X} \in \mathrm{Part} (X) }V^{\mathcal{X}}$.

The expression (\ref{eq:e^V_general}) applies to a general state. If we wish
to specify to a Gaussian state $\mathbb{E}_{g}$ then the expression specializes further as
\begin{gather*}
\Xi = \mathrm{Part} (V)^X \, \mathrm{Pair}(g)_X .
%\label{eq:Xi_Feynman_expansion}
\end{gather*}
A similar argument gives the moments.
\begin{Theorem}
Let $\mathbb{E}$ the probability state given by~\eqref{eq:relative_state}, then its moments are given by
\begin{gather*}
G_X = \frac{1}{\Xi} \mathrm{Part} (V)^Y \, \mathrm{Pair}(g)_{Y+X} .
%\label{eq:G_X_Feynman_expansion}
\end{gather*}
\end{Theorem}

\subsection{Calculus for fields}

If a generating function $Z_{G}(j) $ is (Fr\'{e}chet) differentiable about $j=0$ to all orders, then we may work out the components $G(X)$ according to
\begin{gather*}
G(X)=\bigg\{ \prod_{x\in X}\frac{\delta \;\;\;\;\;\;\;}{\delta j ( x ) }\bigg\} Z_{G}(j) \bigg| _{j=0}.
%\label{eq:Z_diff_G}
\end{gather*}

\begin{Notation}
A useful shorthand is to introduce the multiple Fr\'{e}chet differential operator
\begin{gather*}
\frac{\delta \;\;\;}{\delta j^{X}}=\prod_{x\in X}\frac{\delta \;\;\;\;\;\;\;}{\delta j(x) }
\end{gather*}
along with the derivative $\frac{\delta Z_{G} }{\delta j}$ defined by $\frac{\delta Z_{G}(j) }{\delta j}\colon X\mapsto \frac{\delta Z_{G} (j) }{\delta j^{X}}$.
\end{Notation}

In particular, $\frac{\delta Z_{G} }{\delta j}\big| _{j=0}=G$.
We note that
\begin{align*}
 \frac{\delta Z_{G}(j) }{\delta j(x) }& =\frac{\delta }{\delta j(x) }\sum_{n=0}^{\infty }\frac{1}{n!}\int
G_{n}( x_{1},\dots ,x_{n}) j( x_{1}) \cdots j(x_{n}) \,{\rm d}x_{1}\cdots {\rm d}x_{n} \\
&= \sum_{n=1}^{\infty }\frac{1}{(n-1)!}\int G_{n} ( x,x_{2},\dots,x_{n}) j ( x_{2} ) \cdots j ( x_{n} )
\,{\rm d}x_{2}\cdots {\rm d}x_{n} \\
&= \int G ( \{x\}+Y ) j^{Y}\,{\rm d}Y.
\end{align*}
This may be written more succinctly as $\frac{\delta }{\delta j(x) } G_Y j^Y = G_{Y+\{x\} } j^Y$. More generally we have
\begin{gather*}
\frac{\delta Z_{G}(j) }{\delta j^{X}}=\int G ( X+Y ) j^{Y}\,{\rm d}Y ,
\end{gather*}
or $\frac{\delta Z_{G}(j) }{\delta j} =\int G(\cdot +Y)j^{Y}\,{\rm d}Y$.

\begin{Proposition}[the Leibniz rule for fields] For analytic functionals $U$ and $V$ we have that
\begin{gather*}
\frac{\delta }{\delta j} ( U V ) =\frac{\delta U }{\delta j}\diamond \frac{\delta V }{\delta j}.
\end{gather*}
\end{Proposition}

To see this, set $U=Z_{F}$ and $V=Z_{G}$, then we are required to show that
\begin{gather*}
\frac{\delta }{\delta j^{X}}Z_{F}(j) Z_{G}(j)
=\sum_{X_{1}+X_{2}=X}\frac{\delta Z_{F}(j) }{\delta j^{X_{1}}}
\frac{\delta Z_{G}(j) }{\delta j^{X_{2}}}.
\end{gather*}
The proof follows by elementary induction. For several terms we just find
multiple Wick products.

We can also investigate the form of the chain rule. For a given functional $Z=Z(j) $,
\begin{gather*}
\frac{\delta h(Z) }{\delta j(x_{1}) } =h^{\prime}(Z) \frac{\delta Z}{\delta j(x_{1}) }, \\
\frac{\delta ^{2}h(Z) }{\delta j(x_{1}) \delta j(x_{2}) } =h^{\prime \prime }(Z) \frac{\delta Z}{\delta j(x_{1}) }\frac{\delta Z}{\delta j(x_{2}) }
+h^{\prime }(Z) \frac{\delta ^{2}Z}{\delta j(x_{1})\delta j(x_{2})}, \\
\cdots\cdots\cdots\cdots\cdots\cdots\cdots\cdots\cdots\cdots\cdots\cdots\cdots
\end{gather*}
the pattern is easy to spot and we establish it in the next lemma.

\begin{Lemma}[chain rule for fields]\label{lem:chain rule} Let $Z=Z(j) $ possess Fr\'{e}chet derivatives to all orders and let $h$ be smooth analytic, then
\begin{gather*}
\frac{\delta \;}{\delta j^{X}}h( Z(j) ) =\sum_{\mathcal{X}
\in \mathrm{Part}(X) }h^{( N( \mathcal{X}))}( Z(j)) \left( \frac{\delta Z}{\delta j}\right)_{\mathcal{X}},
\end{gather*}
where $h^{(n) }(\cdot) $ is the $n$th derivative of $h$.
\end{Lemma}

\begin{proof}This is easily seen by induction on. As $\frac{\delta h(Z) }{\delta j^{x}}=h^{\prime }(Z) \frac{\delta Z}{\delta j(x) }$,
the identity is trivially true for $n=1$. Now assume that it is true for $n$, and let $|X|=n$, then
\begin{align*}
\dfrac{\delta \;}{\delta j(x) }\frac{\delta h(Z) }{\delta j^{X}} ={} &\dfrac{\delta }{\delta j(x) }\sum_{\mathcal{X} \in
\mathrm{Part}(X) }h^{(N(\mathcal{X})) }( Z(j)) \left( \frac{\delta Z}{\delta j}\right) _{\mathcal{X}} \\
={} &\sum_{\mathcal{X} \in \mathrm{Part}(X) }h^{( N(\mathcal{X})
+1) }(Z(j)) \frac{\delta Z}{\delta j (x) }\left( \frac{\delta Z}{\delta j}\right) _{\mathcal{X}} \\
&{} +\sum_{\mathcal{X} \in \mathrm{Part}(X) }h^{( N(\mathcal{X})) }(Z(j))\frac{\delta \;}{\delta j(x)}\left( \frac{\delta Z}{\delta j}\right) _{\mathcal{X}}.
\end{align*}
However, the first term on the right hand side is a sum over all parts of $X+ \{x\} $ having $x$ occurring as a singleton, while the second
term, when differentiated with respect to~$j(x) $, will be a sum over all parts of $X+ \{x\} $ having $x$ in some part containing at least
one element of~$X$. Thus we may write the above as
\begin{gather*}
\frac{\delta h(Z) \;}{\delta j^{X+ \{x\} }}
=\sum_{\mathcal{X} \in \mathrm{Part}( X+ \{x\}) }h^{(N(\mathcal{X})) }(Z(j)) \left( \frac{\delta Z}{\delta j}\right) _{\mathcal{X}}.
\end{gather*}
The identity then follows by induction.
\end{proof}

The chain rule is in fact a generalization of the Fa\`{a} di Bruno formula.

\subsection{Feynman diagrams}\label{subsec:FeynDiag}

\begin{Corollary}Given an analytic functional $V[\phi] = \!\int\! v^{X}\phi _{X}{\rm d}X$, where we assume that
\mbox{$V[0] =0$}, then
\begin{gather}
\mathbb{E}\big[ {\rm e}^{V( \phi ) ] }\big] = \bigg( \sum_{\mathcal{X} \in \mathrm{Part}(X) }v_{\mathcal{X}}\bigg) G_{X}, \label{eq:e^V_general}
\end{gather}
where $v_{\mathcal{X}}\triangleq \prod_{A\in \mathcal{X} }v^{A}$.
\end{Corollary}

\begin{proof}
We have $v^{\varnothing }=0$, and so we use (\ref{eq:diamond formula}) to get
\begin{gather*}
\mathbb{E} [ V( \phi ) ^{n} ] = ( v^{\diamond
n}) ^{X}G_{X} \equiv n!\bigg( \sum_{\mathcal{X} \in \mathrm{Part}_{n}(X)
}v_{\mathcal{X}}\bigg) G_{X}.
\end{gather*}
The relation (\ref{eq:e^V_general}) then follows by summing the exponential series.
\end{proof}

The expression~(\ref{eq:e^V_general}) applies to a general state. If we wish
to specify to a Gaussian state $\mathbb{E}_{g}$ then the expression specializes further.

\begin{Theorem}Let $\Xi =\mathbb{E}_{g}\big[ {\rm e}^{V( \phi ) }\big] $, then
\begin{gather*}
\Xi \equiv \int \sum_{\mathcal{X} \in \mathrm{Part}(X) }\sum_{\mathcal{P}\in \mathrm{Pair}(X) }v_{\mathcal{X}}g_{\mathcal{P}}\,{\rm d}X.
\end{gather*}
\end{Theorem}

The proof is then just a simple substitution of the explicit form~(\ref{eq:Gaussian_moments}) for the Gaussian moments into~(\ref{eq:e^V_general}).
To understand this expression, let us look at a typical term appearing on the right hand side.

\begin{figure}[h]\centering
\includegraphics[width=0.750\textwidth]{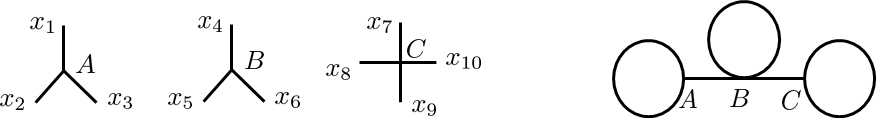}
\caption{With each of the three parts $A$, $B$, $C$ in the partition, we associate a vertex with one line for each of the elements for each part. We then connect the lines to form edges of a graph.}	\label{fig:Feynman_diagram}
\end{figure}

Let us fix a set $X$, say $X=\{ x_{1},\dots
,x_{10}\} $ -- there must be an even number of elements otherwise the
contribution vanishes! We fix a partition $\mathcal{X} =\{ A,B,C\} $ of $X $, say $A=\{ x_{1},x_{2},x_{3}\} $, $B=\{ x_4 ,
x_{5},x_{6}\} $ and $C=\{ x_{7}x_{8},x_{9},x_{10}\} $, and
a pair partition $\mathcal{P}$ consisting of the pairs $(x_{1},x_{2})$, $( x_{3},x_{8})$, $( x_{7},x_{10})$, $( x_{9},x_{5})$, $( x_{4},x_{6})$. The contribution
is the
\begin{gather*}
v^{x_{1}x_{2}x_{3}}v^{x_4, x_{5}x_{6}}v^{x_{7}x_{8}x_{9}x_{10}}g_{x_{1}x_{2}}g_{x_{3}x_{8}}g_{x_{7}x_{10}}g_{x_{9}x_{5}}g_{x_{4}x_{6}},
\end{gather*}
where we have an implied integration over repeated dummy indices. With this contribution we associate the diagram in Figure~\ref{fig:Feynman_diagram}.

More generally we have the following~\cite{GoughKupsch}.

\begin{Theorem}The moments of the state \eqref{eq:relative_state} are given by
\begin{gather*}
G_{X}\equiv \frac{1}{\Xi }\int \sum_{\mathcal{X} \in \mathrm{Part}(Y)
}\sum_{\mathcal{P}\in \mathrm{Pair}( X+Y) }v_{\mathcal{X}}g_{\mathcal{P}}\,{\rm d}Y. %\label{eq:moments_general_expansion}
\end{gather*}
\end{Theorem}

Here the rules are as follows: choose all possible subsets $Y$, all possible
partitions $\mathcal{X} $ of $Y$ and all possible pair partitions $\mathcal{P}$ of $%
X+Y$; draw an $m$-vertex for each part of $\mathcal{X} $ of size $m$, label all the
edges at each vertex by the corresponding elements of $Y$, connect up all
elements of $X+Y$ according the pair partition. We integrate over all $Y$'s,
and sum over all $\mathcal{X} $'s and $\mathcal{P}$'s.

\section{The Dyson--Schwinger equation}

\begin{Definition}Let $g$ be a metric. We define the following operations on function over $\mathscr{P} (M)$ by
\begin{gather*}
(b_x G)_X \triangleq G_{X+\{ x \} } , \\
(b_x^\ast G)_X \triangleq \sum_{x' \in X} g_{xx'} G_{X-\{ x \} } , \\
b^x \triangleq g^{xy} b_y .
\end{gather*}
We also use the notation $b_X = \prod_{x \in X} b_x$, etc.
\end{Definition}

The operators satisfy the canonical commutation relations
\begin{gather*}
[b_x , b_y^\ast ] = g_{xy} , \qquad [b^x , b_y^\ast ] = \delta^x_y .
\end{gather*}
We also note the identity
\begin{gather*}
(b_x G)_X j^X \equiv \frac{\delta }{\delta j^x} \big(G_X j^X \big) .
\end{gather*}

The next result is very easily established.
\begin{Proposition}
The Gaussian functional $Z_g (j) = G_X^g j^X = {\rm e}^{ \frac{1}{2} j^x g_{xy} j^y }$ satisfies
\begin{gather*}
\left(g^{xy} \frac{\delta}{\delta j^y } - j^x \right) Z_g =0 .
\end{gather*}
\end{Proposition}
Alternatively we may look at what this means for the Gaussian moments. We have $ (b_x G^g)_X = G^g_{X + \{ x \} }$.
This will clearly vanish for $X$ even, while for $X$ odd we have that $G^g_{X + \{ x \} }$ will have one factor of the form $g_{xx'}$ for some $x' \in X$ with the other factors being metric terms corresponding to pairs from $X - \{ x' \}$. Therefore, $(b_x G^g)_X = \sum_{x' \in X} g_{xx'} G^g_{X - \{ x' \} } \equiv (b^\ast_x G^g)_X $. This leads to the striking formula for Gaussian moments
\begin{gather*}
 (b_x - b^\ast_x ) G ^g \equiv 0.
\end{gather*}

\begin{Theorem}[the Dyson--Schwinger equation]
Let $\mathbb{E}_G$ be the state, then
\begin{gather*}
\left( g^{xy} \frac{\delta}{\delta j^y } + F_I^x \left( \frac{\delta}{\delta j}\right) - j^x \right) Z_G =0 ,
\end{gather*}
where $F^x_I (\varphi ) = - \frac{ \delta V}{\delta \varphi_x}$. In terms of the moments, we have
\begin{gather*}
\bigg( b_x - b_x^\ast + g_{xy} V^{Y + \{ y \} } b_Y \bigg) G \equiv 0.\label{eq:DSeqt}
\end{gather*}
\end{Theorem}

\begin{proof}The moment generating function $Z_G ( j)$ may be written as
\begin{gather*}
Z_G(j)=\frac{1}{\Xi }\mathbb{E}_{g}\big[ {\rm e}^{ \varphi_x j^x +V[\varphi] }\big] =\frac{1}{\Xi }\exp \left\{ V\left(\frac{\delta }{\delta j} \right) \right\} Z_{g} ( j ) .
%\label{eq:WE}
\end{gather*}
We observe that $
j^{x}Z_G (j) =\frac{1}{\Xi }j^{x}\exp \big\{ V\big( \frac{\delta}{\delta j} \big) \big\} Z_{g} ( j)$,
and using the commutation identity
\begin{gather*}
\big[ j^{x}, {\rm e}^{ V ( \frac{\delta }{\delta j} ) } \big] = F^{x}_I \left( \frac{\delta }{\delta j} \right) {\rm e}^{ V( \frac{\delta }{\delta j} )}
\end{gather*}
we find
\begin{align*}
j^{x}Z_G ( j) &= \frac{1}{\Xi } {\rm e}^{ V ( \frac{\delta }{\delta j} )} j^{x}Z_{g} ( j) +\frac{1}{\Xi }F^{x}_I \left( \frac{\delta }{\delta j} \right) {\rm e}^{ V(\frac{\delta }{\delta j} ) } Z_{g} ( j) \\
&= g^{xy}\frac{\delta }{\delta j^{y}}Z_G(j) +F_I^{x} \left(\frac{\delta }{\delta j} \right) Z_G ( j),
\end{align*}
which gives the result.
\end{proof}

An alternative form of the Dyson--Schwinger equation is
\begin{gather*}
\left\{ F^{x}\left( \frac{\delta }{\delta j}\right) +j^{x}\right\} Z (j) =0 ,
%\label{eq:DS}
\end{gather*}
where $F^{x}(\varphi ) =\frac{\delta S ( \varphi ) }{\delta \varphi _{x}}=-\frac{1}{2}g^{xy}\varphi _{y}+F_{I}^{x} ( \varphi ) $.

The algebraic equations (\ref{eq:DSeqt}) may be written as
\begin{gather}
G_{X+x}=\sum_{x^{\prime }\in X}g_{xx^{\prime }}G_{X-x^{\prime }}+g_{xy}\int_{ \mathscr{P} (M)}
v^{y+Y} G_{X+Y}\, {\rm d}Y.
 \label{eq:DS Green's function}
\end{gather}
We remark that the first term on the right hand side of (\ref{eq:DS Green's function}) contains the moments $\mathbb{E}[ \phi _{X-x^{\prime }}] $ which are of order two smaller than the left hand side $\mathbb{E} [ \phi _{X+x} ] $. The second term on the right hand side of~(\ref{eq:DS Green's function}) contains the moments of higher order, and so we
generally cannot use this equation recursively. In the Gaussian case we have $\mathbb{E}_{g} [ \phi _{X+x} ] =\sum_{x^{\prime
}\in X}g_{xx^{\prime }}\mathbb{E}_{g}[ \phi _{X-x^{\prime }}]
\mathbb{E}_{g}[ \phi _{X-x^{\prime }}] $ from which we can deduce~(\ref{eq:Gaussian field moments}) by just knowing the first and second moments, $\mathbb{E}_{g}[\phi _{x}] =0$ and $\mathbb{E}_{g}[ \phi _{x}\phi _{y}]=g_{xy}$.

\section{Tree expansions}

The generating functions $ Z(j) =\frac{1}{\Xi }\int_{\Phi }{\rm e}^{\langle \varphi
,j\rangle +S[\varphi] } \mathcal{D}\varphi $ may be given a stationary phase approximation
\begin{gather*}
Z ( j ) \simeq {\rm e}^{\psi_x j^x +S (\psi ) }.
\end{gather*}
Assuming that the stationary solution $\psi =\psi ( j) $ exists and is unique for each fixed~$j$, we have the identity
\begin{gather*}
j^{x}+\left. \frac{\delta S( \varphi ) }{\delta \varphi _{x}} \right| _{\varphi =\psi (j ) }=0,
\end{gather*}
that is, $j^{x}-g^{xy}\psi _{y}+\int v^{x+X}\psi _{X}\,{\rm d}X=0$, or rearranging gives (lowering indices using the metric~$g$)
\begin{gather}
\psi _{x}=j_{x}+v_{x}^{\;X}\psi _{X}.\label{eq:classical_expansion}
\end{gather}
We may rewrite (\ref{eq:classical_expansion}) as $\psi_x =j_x + \mathcal{V}_x ( \psi )$ where we have the linear operator $\mathcal{V}_x ( \psi )= v_{x}^{\;X} \psi_X$. Therefore, $( I- \mathcal{V}) \psi = j$ and formally this may be solved as the geometric series $\psi = \sum_{n=0}^\infty \mathcal{V}^n j$. Here, for instance, the second iterate would be
$ (\mathcal{V}^2 j)_x = v_{x}^{\;X} \prod_{x' \in X} \big( v_{x'}^{\;X (x') } j_{X(x')} \big)$. This last expression has an implied Guichardet integration over $X$ and also the $\# X$ variables $X(x')$ labeled by $x' \in X$.

In general, we encounter coefficients of the form $v_{\mathsf{X}_{0}}^{\;\mathsf{X}_{1}} v_{\mathsf{X}_{1}}^{\;\mathsf{X}_{2}} \cdots v_{\mathsf{X}_{n-1}}^{\;\mathsf{X}_{n}}$ in the $n$th iterate where $( \mathsf{X}_0 , \mathsf{X}_1 ,\dots, \mathsf{X}_n )$ is a sequence with $\mathsf{X}_m \in \mathscr{P}^m (M)$.

We may write the expansion in terms of hierarchies. A hierarchy over a finite set $X$ is a~directed tree having subsets of $X$ assigned to each node with the property that if $B \subset X$ is assigned to a particular node then subsets assigned to the immediate daughter nodes form a~partition of~$B$. The hierarchies $\mathrm{Hier}(X)$ are equivalent to the rooted phylogenetic trees whose leaves are the elements of~$X$, as well as the total partitions of~$X$: these are enumerated as sequence A000311 on the Online Encyclopedia of Integer Sequences~\cite{OEIS}. In fact we can give the analytic expansion of~$\psi_x$ in terms of the current~$j$ which in this case is a tree-expansion~\cite{GoughKupsch}
\begin{gather*}
\psi _{x}= \sum_{H \in \text{Hier}(X) } v_{x}(H)j^{X},
\end{gather*}
where the weight $v_{x}(H) $ is calculated by drawing out the tree with $x$ as root and having a factor $ v_{x_B}^{\; x_{A_1} \cdots x_{A_m} }$ for each node where the labels $x_B, x_{A_1},\dots, x_{A_m}$ are dummy variables in~$M$ for the node~$B$ and the daughters $A_1 ,\dots, A_m$.
Apart from the root ($x_B =x$), and the leaves, all these variables are contracted over in the product of such factors (the contractions corresponding to the branches between nodes!).

\subsection*{Acknowledgements}
I would like to thank the anonymous referee who pointed out an issue with the original text presented here as Remark~\ref{prop:ca}.

\pdfbookmark[1]{References}{ref}
\LastPageEnding

\end{document}